\documentclass[aps,prb,twocolumn,showpacs,floatfix]{revtex4-1}
\usepackage{amsmath}
\usepackage{graphicx}
\begin{document}
\title{Rotation-limited growth of three dimensional body-centered cubic crystals}
\author{Jens Tarp and Joachim Mathiesen}
 \affiliation{Niels Bohr Institute, University of Copenhagen, Blegdamsvej 17, DK-2100 Copenhagen, Denmark. }

\begin{abstract}
According to classical grain growth laws, grain growth is driven by the minimization of surface energy and will continue until a single grain prevails. These laws do not take into account the lattice anisotropy and the details of the microscopic rearrangement of mass between grains. Here we consider coarsening of body-centered cubic polycrystalline materials in three dimensions using the phase field crystal model. We observe as function of the quenching depth, a cross over between a state where grain rotation halts and the growth stagnates and a state where grains coarsen rapidly by coalescence through rotation and alignment of the lattices of neighboring grains. We show that the grain rotation per volume change of a grain follows a power law with an exponent of $-1.25$. The scaling exponent is consistent with theoretical considerations based on the conservation of dislocations.

\end{abstract}
\maketitle
\date{}

\section*{Introduction}
In polycrystalline materials, the microstructure, given by the grain sizes, shapes and coordination, is an important "control parameter" for macroscopic material properties like the yield stress, conductivity and brittleness. Polycrystalline materials typically form from the nucleation and growth of grains with different lattice orientations in a quenched or annealed melt. If the annealing temperature is sufficiently high, the grain boundaries will be mobile and rearrange in a way that lead to an overall coarsening of the polycrystalline matrix with time. 

In classical models of grain growth, like the Neumann-Mullins model, grain boundary migration is driven by surface energy and the average grain area grows linearly with time \cite{mullins1956two}, $ \langle A \rangle \sim t $. In three dimensions, however, a correpondingly simple universal growth law can not be established\cite{so36627,macpherson2007neumann} because of the increased geometrical complexity. For models whose dynamics are driven by the minimization of a free energy constrained with a continuity equation, general arguments relying on energy dissipation\cite{kohn2002upper} sets the upper bound for the coarsening dynamics in $ D $ dimensions to $ \langle L^{D} \rangle \sim t^{D/2} $, which coincides with the Neumann-Mullins law. Experimentally it has been observed that the grain size growth can be described by a power law with an exponent that depends on a variety of experimental factors such as the annealing temperature \cite{malow1997grain}. It is further found that the grain growth is sublinear, i.e.~the growth in area is described by a power law exponent less than unity\cite{Huang20002017}. Another possible mechanism, which results in grain coarsening, is the coalescence of neighbouring grains. The rotation of crystals might first lead to the lattice alignment of neighboring grains and secondly the elimination of the grain boundaries between them. Highly convoluted grains can be formed in this way. Grain rotation has been observed in experiments\cite{harris1998grain} and has recently also been observed in molecular-dynamics simulations\cite{Haslam2001293}. 

In many materials, the coarsening dynamics further differs from normal grain growth, in the sense that the grain size distribution varies in time. These abnormal grain growth systems are characterized by a minority of "abnormal" grains growing faster than the mean size leading to an inhomogeneous size distribution\cite{Rollett19891227}. In contrast to classical grain growth laws, grain growth are also known to stagnate in time\cite{barmak2013grain}. In two dimensional systems, it has been suggested that the stagnation might be a combination of high kinetic barriers, preventing mass migration across grain boundaries, and a locking of individual grains preventing grain rotation and subsequent lattice alignment\cite{Bjerre2013}.

In this article, we present a numerical study of the influence of grain rotation on the coarsening dynamics in three dimensional polycrystalline systems with a body-centered cubic (BCC) symmetry.

\section*{Model and Analysis}
To study the coarsening dynamics, we use the phase field crystal model (PFC). The PFC model describes the evolution of a continuous order parameter field, which is spatially periodic with atomistic resolution. The model is based on the minimization of a phenomenological Swift-Hohenberg free energy functional given by  
\begin{equation}
\mathcal{F} = \int d\mathbf{r}\left( \frac{1}{2}\psi \left(1 + \nabla^{2}\right)^{2}\psi  + \frac{a_{2}}{2}\psi^{2} + \frac{\psi^{4}}{4} \right)\label{fe}
\end{equation}
with the order parameter $\psi$ representing the crystal density field\cite{elder2004modeling}. The parameter $a_{2}$ and the mean density $\bar{\psi}$ are related to the melting temperature according to the phase diagram in Fig.~\ref{phase_diagram}. The points in Fig.~\ref{phase_diagram} show the parameters used in the simulations presented in this article. While we change both the mean density and the parameter $a_2$ in our simulations presented here, we shall for convenience only refer to the $a_2$ value in the figures and in the text. The corresponding values for the mean density can then be found from the phase diagram.  Alternatively the functional form of Eq.~(\ref{fe}) can be derived from classical density functional theory \cite{PhysRevB.75.064107}. In three dimensions a free energy on this form results in a rich phase diagram with multiple equilibrium phases. Here, we shall focus only on the BCC phase. The evolution equation of the density field $\psi$ is assumed to obey an over damped diffusion equation 
\begin{equation}
 \frac{d\psi}{dt} = \nabla^{2}\frac{\delta \mathcal{F}[\psi]}{\delta \psi}
\end{equation}
Thereby establishing a link between the microscopic length scales and diffusive time scales. The PFC model has been used to study a wide range of phenomena including phase transitions\cite{PhysRevB.75.064107}, plastic deformation \cite{PhysRevLett.113.265503} and has been shown to successfully predict grain boundary energy as a function of mis-orientation \cite{elder2004modeling}. We solve the dynamical equation using an exponential time integration scheme\cite{Cox2002} in a three dimensional box of size $L \times L \times L$ with $L=512dx$ or $L=1024dx$. Time and space are discretized with $dx=\frac{\pi}{3}$ and $dt=1/2$. We initialize our system from an undercooled melt by introducing small crystal seeds at random points and with random lattice orientations. After the initial crystallization phase, where the seeds grow to cover the whole melt, a polycrystalline structure is formed which coarsen over time. During the coarsening stage, we track the volume and lattice orientation of all grains.  

For the segmentation of the grains, the peaks of the density field are located by a thresholding procedure from which the coordinates of the center of mass of each peak can be calculated. Grain boundary detection is performed using a Voronoi tessellation and a centro symmetry parameter \cite{PhysRevB.58.11085}. The angle and axis of rotation for each grain is found by taking the mean over the orientation of all the unit cells in the interior of the grain. The misorientation angle of two grains $A$ and $B$ is calculated by constructing the rotation matrix $G$ for the grains and form the product $\Delta G_{AB} = G_{A}G^{-1}_{B}$. The misorientation angle $\theta$ is then given by $\theta = \arccos \left( \frac{\text{Tr}(\Delta G_{AB}) - 1 }{2} \right)$.

To get a measure of the typical grain size in our systems, we calculate a coarsening parameter by fitting a Lorentzian squared to the structure factor of the system\cite{PhysRevLett.75.2152}, $\langle |F[\psi]|^{2} \rangle$. Where $F[\psi]$ is the Fourier transform and the outer brackets denote averaging over all orientations in $k$-space. The width $\xi^{-1}(t)$ of the averaged spectrum provides a measure of the ordering scale of the system, which is inversely proportional to the mean grain size. 
\begin{figure}[htbp]
  \includegraphics[width=.47\textwidth]{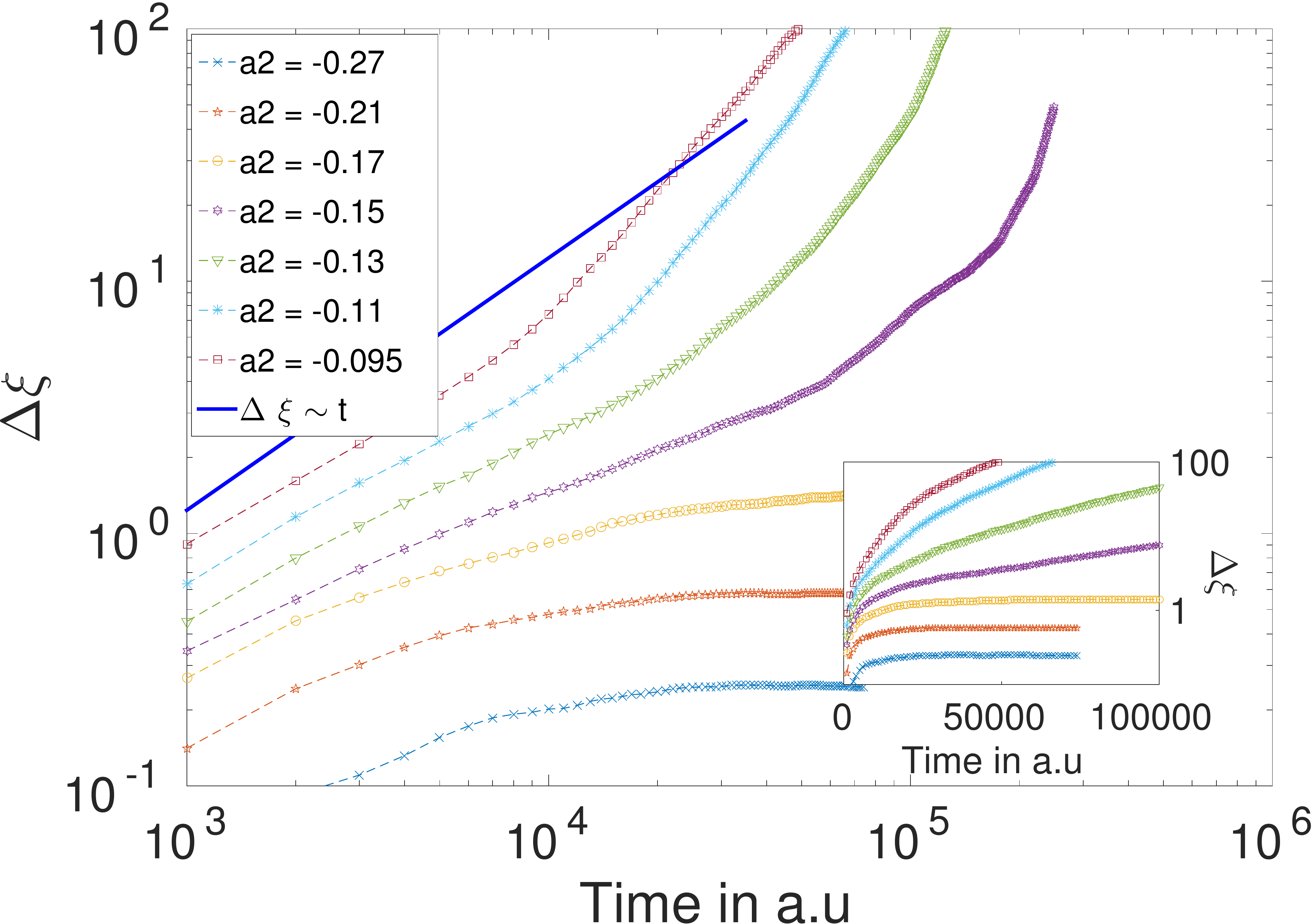}
  \caption{Mean size of grains on double logarithmic axes. The inset shows the same data on semi logarithmic axes.}
  \label{grain_size}
\end{figure} 

\section*{Results}
\begin{figure}[htbp]
	\includegraphics[width=.47\textwidth]{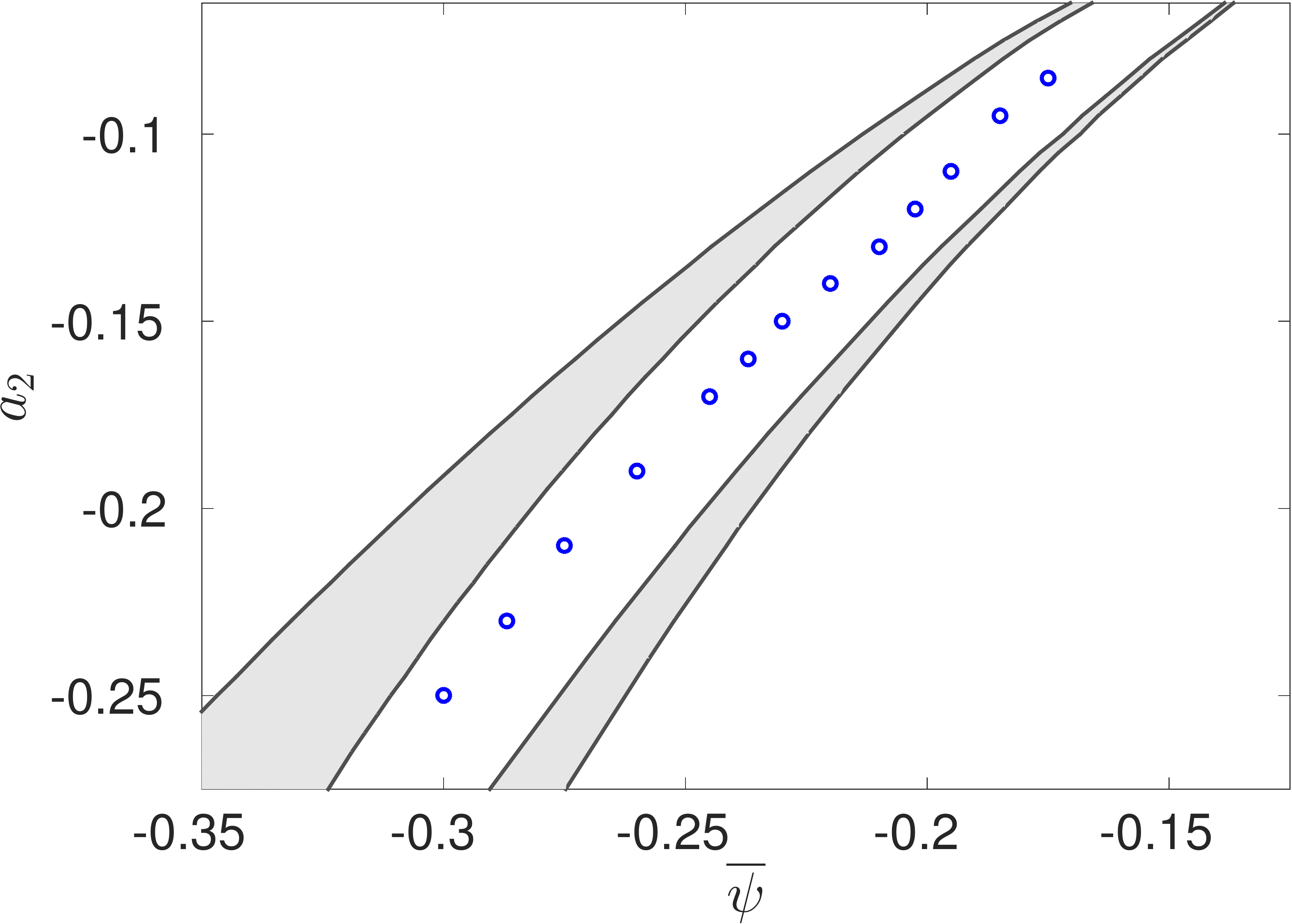}
	\caption{The PFC phase diagram with points indicating the parameters used in the simulations. The region in the middle has a BCC lattice as its equilibrium state while the equilibrium states of the regions to the left and right are given by a uniform liquid state and a rod state, respectively. The gray areas indicates the coexistence regions.}
	\label{phase_diagram}
\end{figure}

\begin{figure}
  \includegraphics[width=.47\textwidth]{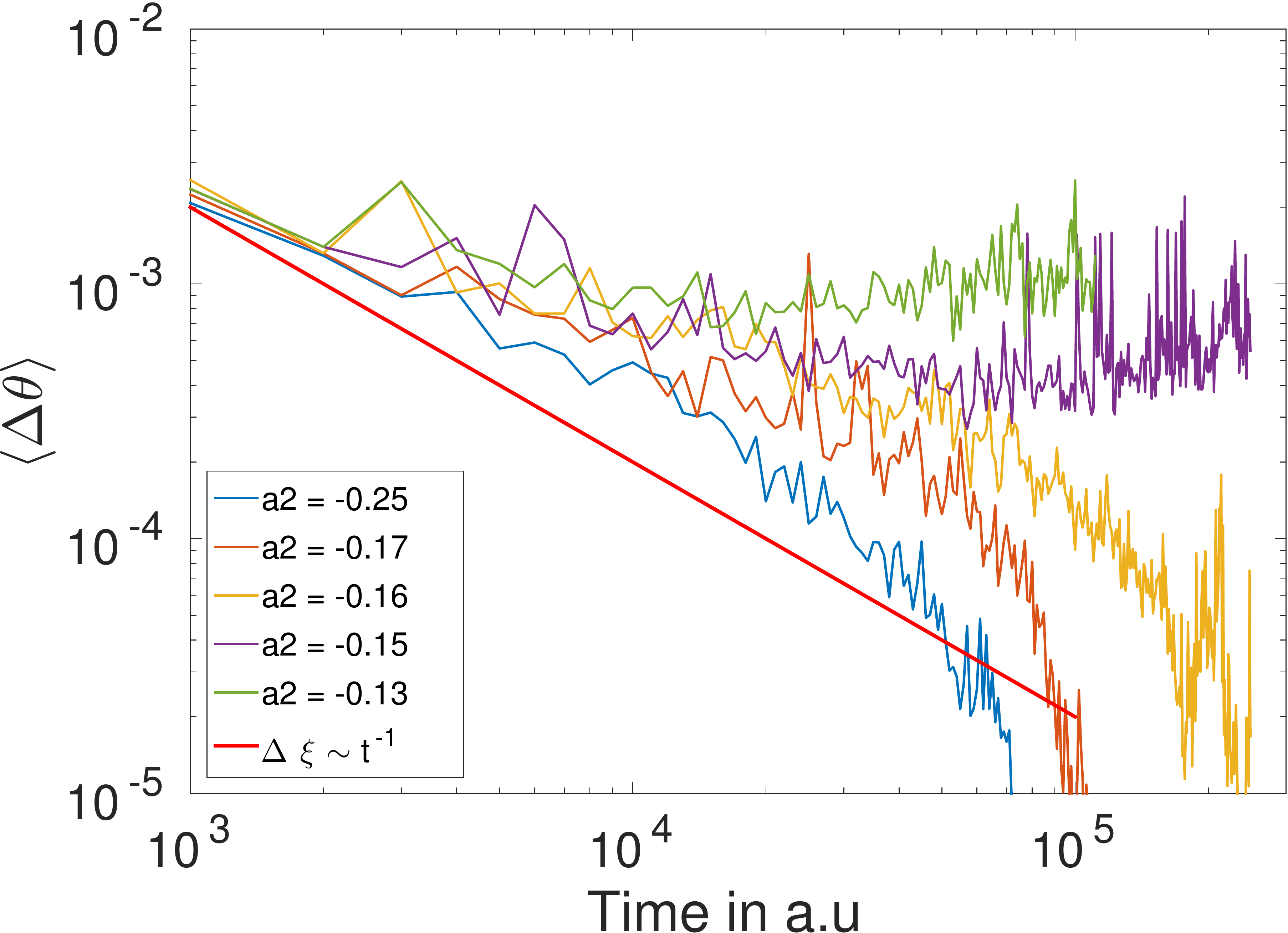}
  \caption{Average rotational change as function of time. The straight line corresponds to a power law with an exponent $n=-1$. The stagnating regime shows a rapid decrease in grain rotation in contrast to the fast coarsening regime where grain rotation continues.}
  \label{delta_T}
\end{figure}

In general, the polycrystalline structure coarsen at a rate which depend on the quenching depth, the $a_2$ parameter in Eq.~(\ref{fe}). In Fig.~\ref{grain_size}, we show the change in the ordering length, $\Delta \xi = \xi(t) - \xi_{i}$, as a function of time for different quenching depths. Where $\xi_{i}$ is the initial ordering scale measured right after the system has fully crystallized. For deep quenching parameters, the system dynamics is described by a short period of grain growth followed by stagnation. As the quenching depth is decreased, the time it takes to reach the stagnation stage is increased and eventually (for $a_2\gtrsim -0.15$ and $\bar \psi\gtrsim -0.22$ in the phase diagram, Fig.~\ref{phase_diagram}) the stagnation stage is replaced by a stage of rapid grain growth. By tracking the orientation of individual grains, we observe that the stagnation is accompanied by a general decrease in the change of the lattice orientation. The change in orientation eventually drops to zero when the system stagnates, see Fig.~\ref{delta_T}.

Independent of the system size, the average grain size, here represented by $\Delta \xi$, can in the early state be approximated by a power law in time, $ \langle \Delta \xi  \rangle \sim t^n $ with an exponent $n$ that depends on the quenching parameters $ a_{2} $ and $\bar \psi$. From simulations, we find values of the exponent in the interval $ n \sim 0.5 -1 $. 
In order to estimate the variation of the coarsening exponent, we have performed repeated simulations for the same quenching depths and we find a standard deviation in the power-law exponents of the order of $0.1$.

\begin{figure}[!htbp]
  \includegraphics[trim = 0mm 0mm  0mm 0mm, clip,width=.47\textwidth]{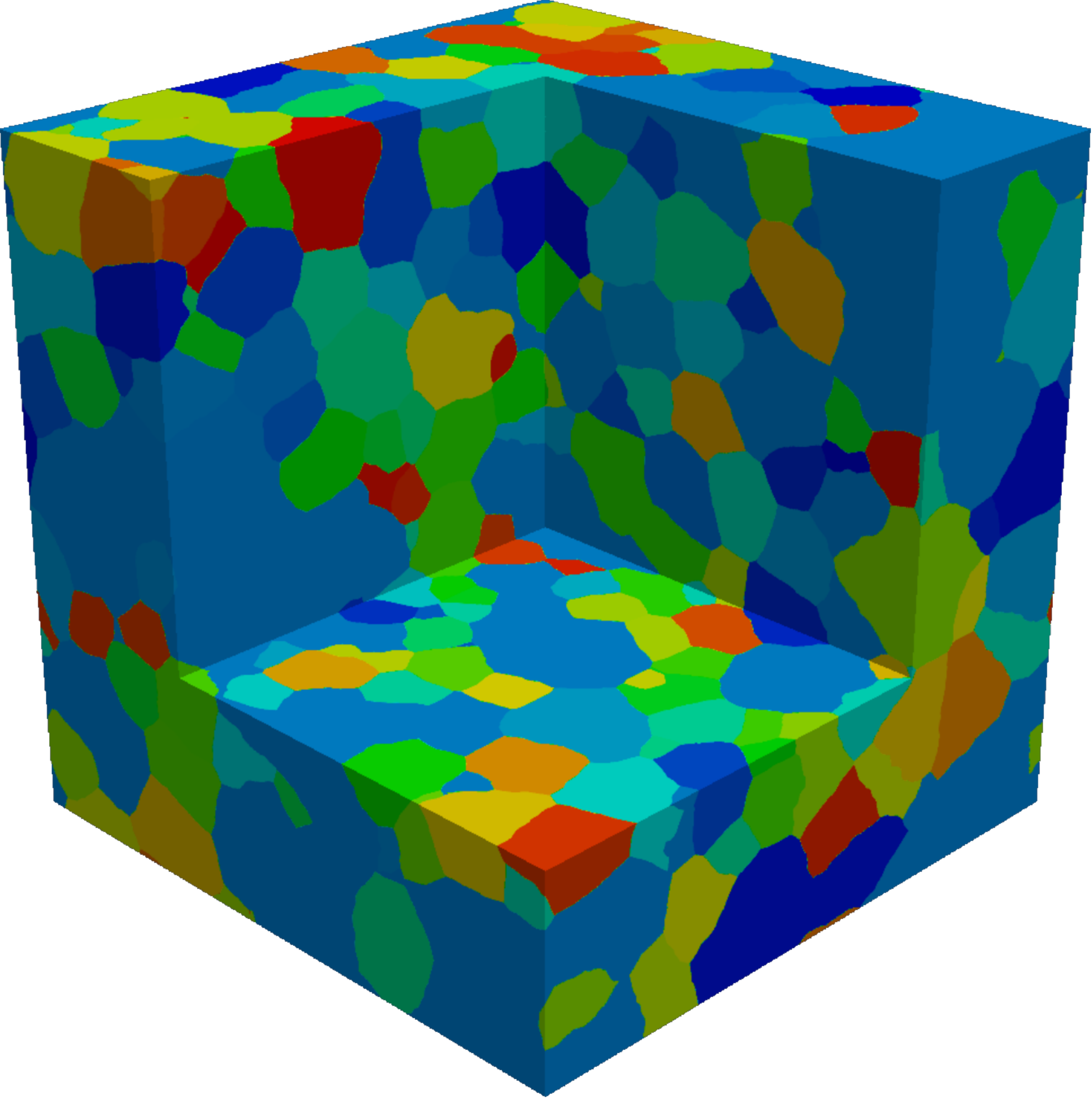}
  \caption{Snapshot of a $L = 1024 dx$ simulation performed with $a_{2}=-0.15$ and with periodic boundary conditions. Note the heterogeneous grain distribution which can be observed at the transition to abnormal grain growth.}
  \label{cross_section}
  \includegraphics[trim = 0mm 0mm 0mm 0mm, clip,width=.47\textwidth]{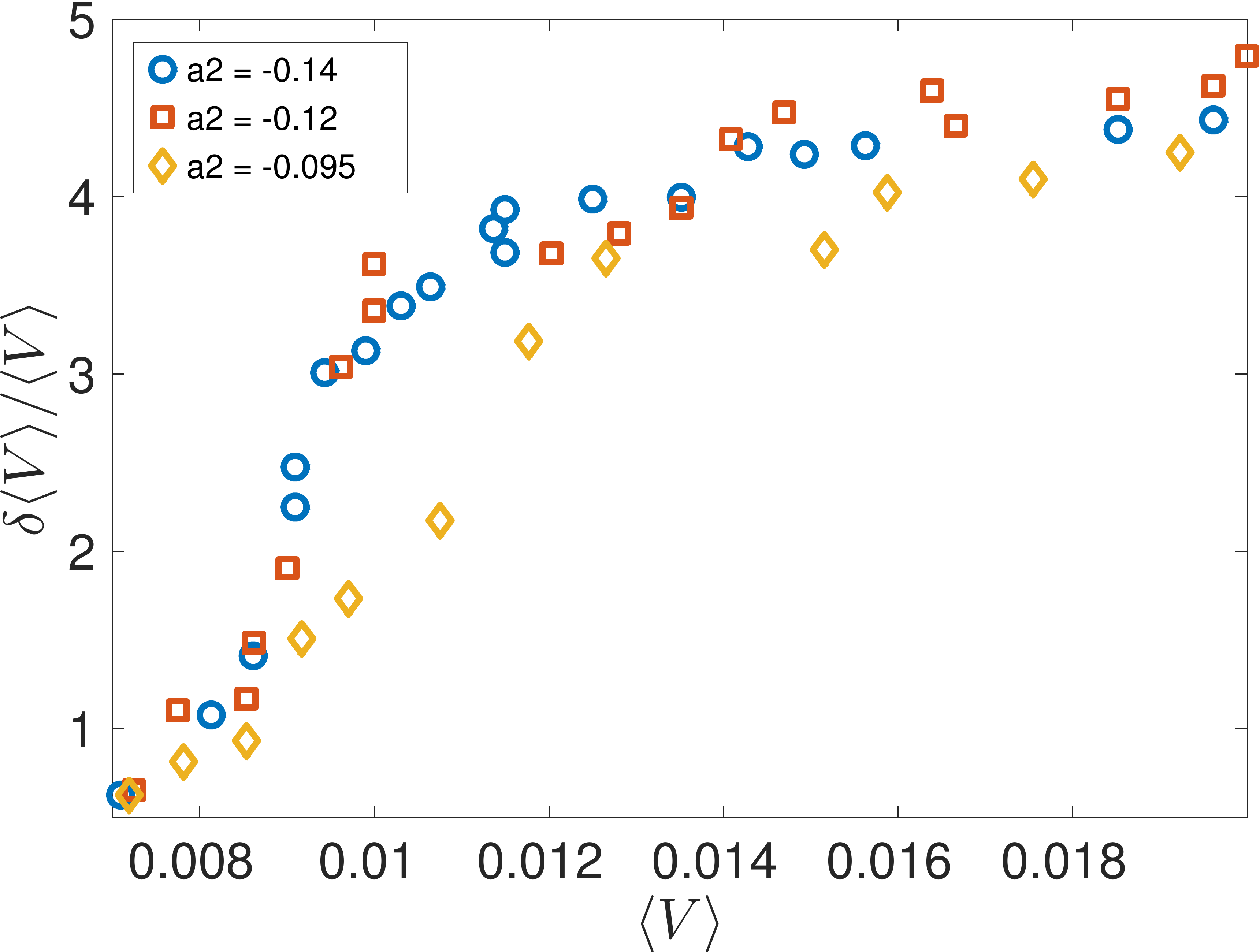}
  \caption{The ratio of the standard deviation in grain size and the mean size as a function of mean size. The non-linear relation is a sign of abnormal grain growth.}
  \label{standard_dev}
\end{figure}

In general, the coarsening dynamics exhibits abnormal grain growth where a few grains grow significantly faster than the rest of the crystal matrix as can be seen in Fig.~\ref{cross_section}. The figure shows a late stage snapshot of a $L = 1024 dx$ system. A consequence of abnormal growth is that the grain size distribution is not self-similar as can be seen from Fig.~\ref{standard_dev}, where the ratio of the standard deviation of the grain sizes and the mean grain volume is plotted. It is also clear from this figure that there is a strong dependence of the rotation on the parameter $a_{2}$. 

\begin{figure}[!htbp]
  \includegraphics[trim = 0mm 0mm 0mm 0mm, clip,width=.47\textwidth]{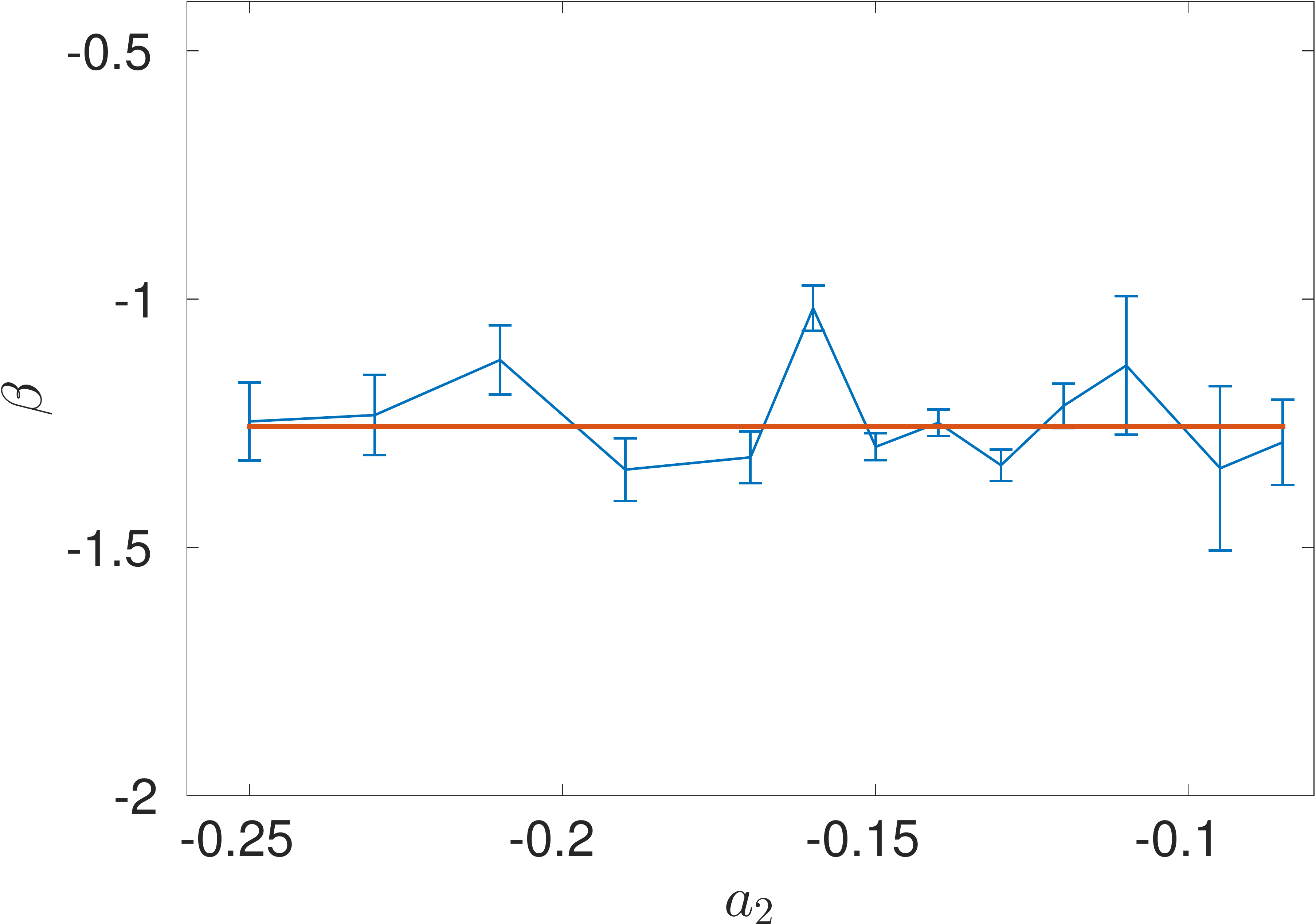}
  \caption{Powerlaw coefficients obtained from best fits to scatter plots of $| {\Delta \theta/}{\Delta V} |$ vs $V$ for different values of $a_{2}$. The uncertainty is the standard deviation calculated using a bootstrap method.}
  \label{theta_power}
\end{figure}

The transition to fast grain growth is initiated by the mobilization of small grains, which align in lattice orientation with the larger so-called abnormal grains and finally coalesce in to an even larger grain. This is supported by the observation that the amount of rotation per change in volume $V$ increases as the small grains get even smaller. From the scatter plots of $|\Delta \theta /\Delta V|$ versus $V$, we fit the general trend by a power law whose exponent is plotted in Fig.~\ref{theta_power}. Best fits across multiple quenching depths and initial conditions suggests a scaling behaviour
\begin{equation}
 \bigg| \frac{\Delta \theta}{\Delta V}\bigg |  \sim V^{-\beta},
\end{equation} 
where the scaling exponent is estimated to $\beta = 1.25 \pm 0.06 $. The uncertainty is taken to be the standard deviation found by fitting a straight line using a weighted least squares approach.
%
%
%

%
%
%
\section*{Discussion}
In our simulations, we observe two distinctly different types of coarsening dynamics. For deep quenches the grain growth is typically accompanied by a decreasing rotation of the individual grains. When the grain rotation stops, the system has reached a stagnation state. Similar results have been reported in two dimensional systems \cite{Holm28052010,Bjerre2013}. For shallow quenches, we find that the early stage coarsening is described by a power law growth of the form $V \propto \Delta \xi \sim t^{n}$ with $n\sim 0.5-1 $, in reasonably agreement with experimental studies\cite{Huang20002017}, which finds $n\sim 0.4 - 1.2$. Further the grain growth is found to be abnormal resulting in a crystal matrix consisting of a few large convoluted grains. Abnormal grain growth has been seen in other numerical experiments \cite{0965-0393-19-4-045008} as well as in experiments\cite{ferry1996discontinuous}. To analyse the late stage coarsening dynamics, we consider a simple model where the grain growth is mediated purely by grain coalescence. The grain coalescence follows from the continuous rotation of predominantly small grains, which tend to align their lattice with larger neighboring grains. Following \cite{Moldovan20023397} we introduce a characteristic time $t_{l}$ it takes for two grains to coalesce. The dynamics of the number of grains can then be described by
\begin{equation}
 \frac{1}{N} \frac{dN}{dt} = - \frac{1}{t_{l}}
\end{equation} 
Since coalescence between two grains happen when their lattices are rotated to be aligned, we assume $t_{l}$ to be proportional to the inverse of the mean grain rotation, $ \frac{1}{\langle \Delta \theta \rangle}$. Using the conservation equation between the total number of grains and mean volume $N(t)\langle V \rangle = V_{sys} $ and differentiating with respect to time and using the coalescence assumption, we establish the following relation
\begin{equation}
 \frac{1}{\langle  V \rangle}\frac{d \langle V \rangle}{dt} = - \frac{1}{N(t)}\frac{dN(t)}{dt} = \langle \Delta \theta \rangle
\end{equation} 
Assuming $\langle \Delta \theta \rangle = C t^{\gamma}$ we find the two solutions
\begin{align}
 \langle V(t) \rangle &= V_{0}\left( \frac{t}{t_{0}}\right)^{C} \qquad \gamma = -1 \\
\langle V(t) \rangle &=V_{0}\exp\left(\frac{C}{(\gamma+1)} \left(t^{\gamma+1} - t_{0}^{\gamma+1} \right)\right)  \qquad
\gamma \neq -1 \label{solutions}
\end{align} 
From these solutions, we see that if the degree of rotation in the system does not fall off quickly enough grain coalescence will lead to  exponential coarsening in time. In Fig.~\ref{delta_T}, we observe that for the late stages of the grain growth, the decay of $\langle \Delta \theta \rangle$ is in general much slower than $t^{-1}$ and consequently, we expect from Eq.~(\ref{solutions}) an exponential grain growth. The fact that we indeed observe an exponential grain growth in the simulations supports the hypothesis that the grain rotation is the primary mediator of grain growth. From Fig.~\ref{delta_T}, we find that in the coarsening regime where $ \gamma \sim 0 $ the proportionality constant $ C $ increases for increasing $ a_{2} $ in agreement with the inset in Fig~\ref{grain_size}.  

We can further estimate the coupling between grain growth and rotation, 
\begin{equation}
\bigg| \frac{\Delta \theta}{\Delta V}\bigg |  \sim V^{-\beta}
\end{equation}
by assuming that the grain rotation is predominantly given by the normal motion of the grain boundary
\begin{equation}
 r(t)\frac{d\theta }{dt} = v_{n}.
\end{equation} 
The conservation of dislocations then implies that\cite{cahn2004unified}  $r(t)\theta(t) = const$, or equivalently, $\theta(t) \sim V^{-1/3}$, from which it follows that $\beta = {4}/{3}$ in reasonably good agreement with our numerical results. The fact that simulations agree with this simple estimate might be an indication that dislocations are indeed conserved up to the point where grains coalesce.

From simulations of grain growth in polycrystalline materials, we have identified two distinct dynamical regimes. One regime for deep quenching parameters where grain rotation quickly is suppressed and therefore leads to an overall stagnation of growth. Another regime is observed for more shallow quenches where grains continue to rotate and therefore will be able to align their lattices. The alignment eventually leads to the coalescence of neighbouring grains and simultaneously allows for a few abnormally large grains to form. The grain rotation and coalescence cause an exponential increase in grain growth with time. 

\textit{Acknowledgments}. This study was supported through the
grant Earth Patterns by the Villum Foundation 

\bibliographystyle{unsrt}
\bibliography{reference}
\end{document}